# Challenges and Opportunities of Big Data in Healthcare Mobile Applications

**Mohsen Aghabozorgi Nafchi[1], Maryam Aghabozorgi Nafchi[2]**

1- Shiraz University- Mohsen.aghabozorgi@gmail.com

2- Shahrekord Uniersity- ma.aghabozorgi@gmail.com

**Abstract**

The health and various ways to improve healthcare systems are one of the most concerns of human in history. By the growth of mobile technology, different mobile applications in the field of the healthcare system are developed. These mobile applications instantly gather and analyze the data of their users to help them in the health area. This volume of data will be a critical problem. Big data in healthcare mobile applications have its challenges and opportunities for the users and developers. Does this amount of gathered data which is increasing day by day can help the human to design new tools in healthcare systems and improve health condition? In this chapter, we will discuss meticulously the challenges and opportunities of big data in the healthcare mobile applications.

## 1-Introduction

By expanding mobile technology, many mobile apps have been developed in many areas. One of these areas is the healthcare[1]. As of 2018, global trade in mobile health app is estimated at the US $ 28.320 billion and this amount will reach the US $ 102.35 billion in 2023 [2]. It is also estimated that in 2019, Digital health tech will grow by 30%, reaching $ 25 billion that One of the reasons for this is the rise in health and aging costs, which makes it possible to pay attention to digital health tech, such as the mobile health app[3]. Mhealth apps can be divided into areas such as education, electronic health record, human resource management, supply chain management, and Vital Events Tracking[4].

On the other hand, today, by collecting and recording data, we are witnessing the development of a new concept called big data, which has three attributes such as varieties, velocity and volume[5] that In the health domain, three other attributes of value, veracity, and variability can also be considered[6]. With developing applications of big data, there is a new era in decision making in different areas[7].

By increasing mhealth apps and recording patient-related data and their vital signs for analysis and diagnosis of diseases, the concept of big data will play a key role in the field of health and will create many challenges and opportunities for the development of new technologies in this area[6][8][9][10]. This study first explores the opportunities and then addresses the challenges ahead in mhealth apps using big data capacities.

## 2- Opportunities with Big Data in Mobile Healthcare

### 2-1- Recommendation Systems in HealthCare mobile applications

With high-level data processing data, recommender systems offer most relevant products to people[11]. recommender systems in mhealth apps are divided into 5 groups[12].

1. **Nutrition:**
   This group of mhealth apps offers users to consume calories and food in areas such as obesity[13][14][15].
2. **Physical Activities:**
   This category of mhealth apps offers physical activities to users[16][17].
3. **Nutrition and Physical activities:**
   In this type of mhealth apps, users are advised to consume food and exercise[18][19]
4. **Several Applicability Areas:**
   This type of mhealth apps uses various tools such as cloud, databases of health information, wireless sensors, user behavior and a prediction matrix to suggest recommendations in health and physiology to users[20].

Due to the large volume of collected data with mobile phones and the existence of big data, this collected data can be used to improve the quality of Recommender Systems in mhealth apps.

### 2-2- Epidemic Surveillance Mobile Health Applications

Epidemiological surveillance systems are used to control and prevent diseases[21]. So, Due to availability of mobile phones, mhealth apps can play an effective role in preventing the spread of diseases around the world[22] and used in epidemiological surveillance systems . In the field of Ebola epidemic control, one of the most important challenges was access to high precision data. For this reason, a smartphone-based contact tracing system[23] has been designed and implemented in Guinea to access data analysis software. On the other hand, they can interact with people who are not visited and monitor them through GPS. In another study in Kenya, in order to diagnose the disease before the outbreak, a mobile-based disease surveillance prototype was used to report and analyze data[24]. Also, It can be seen mhealth apps for Malaria [25][26][27], Influenza[28][29][30][31], HIV [32] and natural disasters[33][34]. So, data collection with mhaealth apps can be an opportunity to analyze and diagnose diseases.

### 2-3- Food Safety Monitoring

In recent years, progresses in sensor technology and connecting them with mobile phones have provided possibility of food safety analyst and drug diagnosis[35]. For example, a sensing smartphone designed to measure marine toxins that uses the data collected to detect the toxin[36]. Also, a tool designed to analyze honey's constituents[37]. Other examples of tools developed alongside the mobile phone are the a microfluidic chip designed to test the taste of red wine[38], Escherichia coli (E. coli) detection in liquids[39], monitoring tetracycline (TC) in milk[40], diagnosis of aflatoxin-B1 in maize[41], detection of BDE-47 in food samples[42], detection of Peanut allergen in food samples[43],and analysis of catechols in water[44]. Big data can play an important role in the development of tools used in the field of food safety.

**2-4- Microbiology and Medical Applications**

One of the important topics in the field of health is the analysis of microbiology data and medical applications. Considering the potential of big data and mobile phones, many opportunities can be seen in the field of microbiology and medical applications[45]. For example, in genetics, a tool have been designed for imaging and sizing of single DNA molecules on smartphones[46].

**2-5- Aging**

Today, One of the important issues is increasing the number of elderly people in societies and one of the main challenges is caring them[47]. Recent developments in the field of mobile health show that mobile devices and data collected by mobile devices can play an important role in improving the quality of healthcare for the elderly[48]. Examples of mhealth apps for ageing are mobile apps for caregivers[49] and sleep[50][51].

**2-6- Air Quality**

One of the important issues in the health issue is air quality, which has a direct impact on human health and big data can play an important role in this area[9]. With the development of various mhealth apps in the field of health, we are witnessing the development of the applications in the field of air quality[52][53][54]. For example, AirRater [55] is a system with environmental monitoring in the field of air pollution and pollen, which allows users to use a mhealth app for getting advice on health such as asthma and allergic rhinitis.

**3-Challenges with Big Data in Mobile Healthcare**

**3-1- Security and Privacy**

Security in big data is divided into four parts: data collection, data storage, data analytics, and knowledge creation[8][56]

5. **Data Collection**

   At this point, the security issues associated with data collection are emphasized. In the mhealth domain, mobile phone security plays an important role in data collection. Health app itself can also be considered a threat because its developers are anonymous and there is no information about how to use the collected data and there are also a variety of attacks such as denial-of-app attack[57]. Bluetooth connection between mobile phone and peripheral devices such as smart watch can jeopardize data transfer security[58]. Phishing and Spoofing attacks are also other threats in the area of data collection[57].

6. **Data Storage**

   In this section, there are threats such as physical stolen data or illegal access to stored data[8][56].

7. **Data Analytics**

   At this point, the use of Data Mining Based Attacks for data analysis threatens patients and users[8][56]. For example, collected GPS data is use in other topics[59]. The use of new data storage techniques can play an important role in preventing such attacks [60].

8. **Knowledge Creation**

   Information extracted from big data and release them could be a serious threat to users of mhealth apps[8][56].

In the area of privacy, collaborators and actors in the field of data mining and big data, they play an important role in promoting the privacy of individuals. data provider, data collector, data miners and decision makers are people who access the data and can threaten the privacy of users[61].

**4- Conclusion**

By using smartphones and installation of different apps, data registration of users is emerged and the concept of dig data is featured more than ever. According to this, data collection from the users by using cell phones and their data analysis has created many opportunities in the healthcare. With the analyzed data, strong applications in the fields of epidemiological surveillance, recommendation systems, food safety monitoring systems, microbiology and medical application, aging and air quality can be designed. On the other, security and privacy are the challenges of data collection in the healthcare.

**References**

[1]   Kiah, M. L. M.; Zaidan, B. B.; Zaidan, A. A.; Nabi, M.; Ibraheem, R. "MIRASS: Medical informatics research activity support system using information mashup network",

*Journal of Medical Systems*, Vol. 38, No. 4, 2014, pp.37

[2]     *Mobile Health (mHealth) App Market - Industry Trends, Opportunities and Forecasts to 2023*

[3]     Behera, K. *Top 8 Healthcare Predictions for 2019*

[4]     Labrique, A. B.; Vasudevan, L.; Kochi, E.; Fabricant, R.; Mehl, G. "mHealth innovations as health system strengthening tools: 12 common applications and a visual framework", *Global Health: Science and Practice*, Vol. 1, No. 2, 2013, pp.160–171

[5]     Laney, D. "3D data management: Controlling data volume, velocity and variety", *META Group Research Note*, Vol. 6, No. 70, 2001, pp.1

[6]     Andreu-Perez, J.; Poon, C. C. Y.; Merrifield, R. D.; Wong, S. T. C.; Yang, G.-Z. "Big data for health", *IEEE J Biomed Health Inform*, Vol. 19, No. 4, 2015, pp.1193–1208

[7]     McAfee, A.; Brynjolfsson, E.; Davenport, T. H.; Patil, D. J.; Barton, D. "Big data: the management revolution", *Harvard Business Review*, Vol. 90, No. 10, 2012, pp.60–68

[8]     Abouelmehdi, K.; Beni-Hessane, A.; Khaloufi, H. "Big healthcare data: preserving security and privacy", *Journal of Big Data*, Vol. 5, No. 1, 2018, pp.1

[9]     Huang, T.; Lan, L.; Fang, X.; An, P.; Min, J.; Wang, F. "Promises and challenges of big data computing in health sciences", *Big Data Research*, Vol. 2, No. 1, 2015, pp.2–11

[10]    Alonso, S. G.; de la Torre Díez, I.; Rodrigues, J. J. P. C.; Hamrioui, S.; López-Coronado, M. "A systematic review of techniques and sources of big data in the healthcare sector", *Journal of Medical Systems*, Vol. 41, No. 11, 2017, pp.183

[11]    Valdez, A. C.; Ziefle, M.; Verbert, K.; Felfernig, A.; Holzinger, A. "Recommender systems for health informatics: state-of-the-art and future perspectives", *Machine Learning for Health Informatics*, Springer, 391–414

[12]    Ferretto, L. R.; Cervi, C. R.; de Marchi, A. C. B. "Recommender systems in mobile apps for health a systematic review", *2017 12th Iberian Conference on Information Systems and Technologies (CISTI)*, IEEE, 1–6

[13]    Li, H.; Zhang, Q.; Lu, K. "Integrating mobile sensing and social network for personalized health-care application", *Proceedings of the 30th Annual ACM Symposium on Applied Computing*, ACM, 527–534

[14]    Jung, H.; Chung, K. "Knowledge-based dietary nutrition recommendation for obese management", *Information Technology and Management*, Vol. 17, No. 1, 2016, pp.29–42

[15]    Ge, M.; Ricci, F.; Massimo, D. "Health-aware food recommender system", *Proceedings of the 9th ACM Conference on Recommender Systems*, ACM, 333–334
5

[16] Vlahu-Gjorgievska, E.; Koceski, S.; Kulev, I.; Trajkovik, V. "Connected-Health Algorithm: Development and Evaluation", *Journal of Medical Systems*, Vol. 40, No. 4, 2016, pp.109

[17] Ho, T. C. T.; Chen, X. "ExerTrek: a portable handheld exercise monitoring, tracking and recommendation system", *2009 11th International Conference on E-Health Networking, Applications and Services (Healthcom)*, IEEE, 84–88

[18] Rabbi, M.; Aung, M. H.; Zhang, M.; Choudhury, T. "MyBehavior: automatic personalized health feedback from user behaviors and preferences using smartphones", *Proceedings of the 2015 ACM International Joint Conference on Pervasive and Ubiquitous Computing*, ACM, 707–718

[19] Wing, C.; Yang, H. "FitYou: integrating health profiles to real-time contextual suggestion", *Proceedings of the 37th International ACM SIGIR Conference on Research & Development in Information Retrieval*, ACM, 1263–1264

[20] Wang, S.-L.; Chen, Y. L.; Kuo, A. M.-H.; Chen, H.-M.; Shiu, Y. S. "Design and evaluation of a cloud-based Mobile Health Information Recommendation system on wireless sensor networks", *Computers & Electrical Engineering*, Vol. 49, 2016, pp.221–235

[21] Ouedraogo, B.; Gaudart, J.; Dufour, J.-C. "How does the cellular phone help in epidemiological surveillance? A review of the scientific literature", *Informatics for Health and Social Care*, Vol. 44, No. 1, 2019, pp.12–30

[22] Wood, C. S.; Thomas, M. R.; Budd, J.; Mashamba-Thompson, T. P.; Herbst, K.; Pillay, D.; Peeling, R. W.; Johnson, A. M.; McKendry, R. A.; Stevens, M. M. "Taking connected mobile-health diagnostics of infectious diseases to the field", *Nature*, Vol. 566, No. 7745, 2019, pp.467

[23] Sacks, J. A.; Zehe, E.; Redick, C.; Bah, A.; Cowger, K.; Camara, M.; Diallo, A.; Gigo, A. N. I.; Dhillon, R. S.; Liu, A. "Introduction of mobile health tools to support Ebola surveillance and contact tracing in Guinea", *Global Health: Science and Practice*, Vol. 3, No. 4, 2015, pp.646–659

[24] Moturi, C. A.; Kinuthia, R. M. "Mobile based Notifiable Disease Surveillance-Case for Kenya", *International Journal of Computer Applications*, Vol. 95, No. 7, 2014

[25] Tatem, A. J.; Huang, Z.; Narib, C.; Kumar, U.; Kandula, D.; Pindolia, D. K.; Smith, D. L.; Cohen, J. M.; Graupe, B.; Uusiku, P. "Integrating rapid risk mapping and mobile phone call record data for strategic malaria elimination planning", *Malaria Journal*, Vol. 13, No. 1, 2014, pp.52

[26] Davis, R. G.; Kamanga, A.; Castillo-Salgado, C.; Chime, N.; Mharakurwa, S.; Shiff, C. "Early detection of malaria foci for targeted interventions in endemic southern Zambia", *Malaria Journal*, Vol. 10, No. 1, 2011, pp.260

[27] Kamanga, A.; Moono, P.; Stresman, G.; Mharakurwa, S.; Shiff, C. "Rural health centres,

communities and malaria case detection in Zambia using mobile telephones: a means to detect potential reservoirs of infection in unstable transmission conditions", *Malaria Journal*, Vol. 9, No. 1, 2010, pp.96

[28] Lin, Y.; Heffernan, C. "Accessible and inexpensive tools for global HPAI surveillance: A mobile-phone based system", *Preventive Veterinary Medicine*, Vol. 98, Nos. 2–3, 2011, pp.209–214

[29] Lajous, M.; Danon, L.; Lopez-Ridaura, R.; Astley, C. M.; Miller, J. C.; Dowell, S. F.; O'Hagan, J. J.; Goldstein, E.; Lipsitch, M. "Mobile messaging as surveillance tool during pandemic (H1N1) 2009, Mexico", *Emerging Infectious Diseases*, Vol. 16, No. 9, 2010, pp.1488

[30] Dia, N.; Sarr, F. D.; Thiam, D.; Sarr, T. F.; Espié, E.; OmarBa, I.; Coly, M.; Niang, M.; Richard, V. "Influenza-like illnesses in Senegal: not only focus on influenza viruses", *PLoS One*, Vol. 9, No. 3, 2014, pp.e93227

[31] Hanafusa, S.; Muhadir, A.; Santoso, H.; Tanaka, K.; Anwar, M.; Sulistyo, E. T.; Hachiya, M. "A surveillance model for human avian influenza with a comprehensive surveillance system for local-priority communicable diseases in South Sulawesi, Indonesia", *Tropical Medicine and Health*, 2012

[32] Laurence, C.; Wispelwey, E.; Flickinger, T. E.; Grabowski, M.; Waldman, A. L.; Plews-Ogan, E.; Debolt, C.; Reynolds, G.; Cohn, W.; Ingersoll, K. "Development of PositiveLinks: A Mobile Phone App to Promote Linkage and Retention in Care for People With HIV", *JMIR Formative Research*, Vol. 3, No. 1, 2019, pp.e11578

[33] Yang, C.; Yang, J.; Luo, X.; Gong, P. "Use of mobile phones in an emergency reporting system for infectious disease surveillance after the Sichuan earthquake in China", *Bulletin of the World Health Organization*, Vol. 87, 2009, pp.619–623

[34] Ma, J.; Zhou, M.; Li, Y.; Guo, Y.; Su, X.; Qi, X.; Ge, H. "Design and application of the emergency response mobile phone-based information system for infectious disease reporting in the Wenchuan earthquake zone", *Journal of Evidence-Based Medicine*, Vol. 2, No. 2, 2009, pp.115–121

[35] Rateni, G.; Dario, P.; Cavallo, F. "Smartphone-based food diagnostic technologies: a review", *Sensors*, Vol. 17, No. 6, 2017, pp.1453

[36] Fang, J.; Qiu, X.; Wan, Z.; Zou, Q.; Su, K.; Hu, N.; Wang, P. "A sensing smartphone and its portable accessory for on-site rapid biochemical detection of marine toxins", *Analytical Methods*, Vol. 8, No. 38, 2016, pp.6895–6902. doi:10.1039/C6AY01384H

[37] Giordano, G. F.; Vicentini, M. B. R.; Murer, R. C.; Augusto, F.; Ferrão, M. F.; Helfer, G. A.; da Costa, A. B.; Gobbi, A. L.; Hantao, L. W.; Lima, R. S. "Point-of-use electroanalytical platform based on homemade potentiostat and smartphone for multivariate data processing", *Electrochimica Acta*, Vol. 219, 2016, pp.170–177

[38] San Park, T.; Baynes, C.; Cho, S.-I.; Yoon, J.-Y. "Paper microfluidics for red wine tasting", *Rsc Advances*, Vol. 4, No. 46, 2014, pp.24356–24362

[39] Zhu, H.; Sikora, U.; Ozcan, A. "Quantum dot enabled detection of Escherichia coli using a cell-phone", *Analyst*, Vol. 137, No. 11, 2012, pp.2541–2544

[40] Masawat, P.; Harfield, A.; Namwong, A. "An iPhone-based digital image colorimeter for detecting tetracycline in milk", *Food Chemistry*, Vol. 184, 2015, pp.23–29

[41] Lee, S.; Kim, G.; Moon, J. "Performance improvement of the one-dot lateral flow immunoassay for aflatoxin B1 by using a smartphone-based reading system", *Sensors*, Vol. 13, No. 4, 2013, pp.5109–5116

[42] Chen, A.; Wang, R.; Bever, C. R. S.; Xing, S.; Hammock, B. D.; Pan, T. "Smartphone-interfaced lab-on-a-chip devices for field-deployable enzyme-linked immunosorbent assay", *Biomicrofluidics*, Vol. 8, No. 6, 2014, pp.64101

[43] Coskun, A. F.; Wong, J.; Khodadadi, D.; Nagi, R.; Tey, A.; Ozcan, A. "A personalized food allergen testing platform on a cellphone", *Lab on a Chip*, Vol. 13, No. 4, 2013, pp.636–640

[44] Wang, Y.; Li, Y.; Bao, X.; Han, J.; Xia, J.; Tian, X.; Ni, L. "A smartphone-based colorimetric reader coupled with a remote server for rapid on-site catechols analysis", *Talanta*, Vol. 160, 2016, pp.194–204

[45] Koydemir, H. C.; Ozcan, A. "Mobile phones create new opportunities for microbiology research and clinical applications", Future Medicine

[46] Wei, Q.; Luo, W.; Chiang, S.; Kappel, T.; Mejia, C.; Tseng, D.; Chan, R. Y. L.; Yan, E.; Qi, H.; Shabbir, F. "Imaging and sizing of single DNA molecules on a mobile phone", *ACS Nano*, Vol. 8, No. 12, 2014, pp.12725–12733

[47] World Health Organization. *Ageing and health*

[48] Cosco, T. D.; Firth, J.; Vahia, I.; Sixsmith, A.; Torous, J. "Mobilizing mHealth Data Collection in Older Adults: Challenges and Opportunities", *JMIR Aging*, Vol. 2, No. 1, 2019, pp.e10019

[49] Grossman, M. R.; Zak, D. K.; Zelinski, E. M. "Mobile Apps for Caregivers of Older Adults: Quantitative Content Analysis", *JMIR MHealth and UHealth*, Vol. 6, No. 7, 2018

[50] Bai, Y.; Xu, B.; Ma, Y.; Sun, G.; Zhao, Y. "Will you have a good sleep tonight?: sleep quality prediction with mobile phone", *Proceedings of the 7th International Conference on Body Area Networks*, ICST (Institute for Computer Sciences, Social-Informatics and …, 124–130

[51] Choi, Y. K.; Demiris, G.; Lin, S.-Y.; Iribarren, S. J.; Landis, C. A.; Thompson, H. J.; McCurry, S. M.; Heitkemper, M. M.; Ward, T. M. "Smartphone applications to support sleep self-management: review and evaluation", *Journal of Clinical Sleep Medicine*, Vol. 14, No. 10,

2018, pp.1783–1790

[52] Re, G. Lo; Peri, D.; Vassallo, S. D. "A mobile application for assessment of air pollution exposure", *Proceedings of the 1st Conference on Mobile and Information Technologies in Medicine (MobileMed 2013)*, Citeseer

[53] Capezzuto, L.; Abbamonte, L.; De Vito, S.; Massera, E.; Formisano, F.; Fattoruso, G.; Di Francia, G.; Buonanno, A. "A maker friendly mobile and social sensing approach to urban air quality monitoring", *SENSORS, 2014 IEEE*, IEEE, 12–16

[54] Leonardi, C.; Cappellotto, A.; Caraviello, M.; Lepri, B.; Antonelli, F. "SecondNose: an air quality mobile crowdsensing system", *Proceedings of the 8th Nordic Conference on Human-Computer Interaction: Fun, Fast, Foundational*, ACM, 1051–1054

[55] Johnston, F. H.; Wheeler, A. J.; Williamson, G. J.; Campbell, S. L.; Jones, P. J.; Koolhof, I. S.; Lucani, C.; Cooling, N. B.; Bowman, D. "Using smartphone technology to reduce health impacts from atmospheric environmental hazards", *Environmental Research Letters*, Vol. 13, No. 4, 2018, pp.44019

[56] Alshboul, Y.; Nepali, R.; Wang, Y. "Big data lifecycle: threats and security model", 2015

[57] Hussain, M.; Zaidan, A. A.; Zidan, B. B.; Iqbal, S.; Ahmed, M. M.; Albahri, O. S.; Albahri, A. S. "Conceptual framework for the security of mobile health applications on android platform", *Telematics and Informatics*, Vol. 35, No. 5, 2018, pp.1335–1354

[58] Naveed, M.; Zhou, X.; Demetriou, S.; Wang, X.; Gunter, C. A. "Inside Job: Understanding and Mitigating the Threat of External Device Mis-Binding on Android.", *NDSS*

[59] Karim, W. "The privacy implications of personal locators: why you should think twice before voluntarily availing yourself to GPS monitoring", *Wash. UJL & Pol'y*, Vol. 14, 2004, pp.485

[60] Dev, H.; Sen, T.; Basak, M.; Ali, M. E. "An approach to protect the privacy of cloud data from data mining based attacks", *2012 SC Companion: High Performance Computing, Networking Storage and Analysis*, IEEE, 1106–1115

[61] Xu, L.; Jiang, C.; Wang, J.; Yuan, J.; Ren, Y. "Information security in big data: privacy and data mining", *Ieee Access*, Vol. 2, 2014, pp.1149–1176